\documentclass{aa}
\usepackage{graphicx}
\usepackage{txfonts}

\begin{document}

\title{VLA multifrequency observations of RS~CVn binaries}

\author{J. Garc\'{\i}a-S\'anchez
          \and J.~M. Paredes\thanks{CER on Astrophysics, Particle Physics
and Cosmology. Universitat de Barcelona}
          \and M. Rib\'o
}

\offprints{J. Garc\'{\i}a-S\'anchez \\ \email{jgarcia@am.ub.es}}

\institute{Departament d'Astronomia i Meteorologia, Universitat de
Barcelona, Av. Diagonal 647, 08028 Barcelona, Spain\\
\email{jgarcia@am.ub.es; josep@am.ub.es; mribo@am.ub.es}
}

\date{Received / Accepted}

\abstract{
We present multiepoch Very Large Array (VLA) observations at 1.4~GHz, 4.9~GHz,
8.5~GHz and 14.9~GHz for a sample of eight RS~CVn binary systems. Circular
polarization measurements of these systems are also reported. Most of the
fluxes observed are consistent with incoherent emission from mildly
relativistic electrons. Several systems show an increase of the degree of
circular polarization with increasing frequency in the optically thin regime,
in conflict with predictions by gyrosynchrotron models. We observed a reversal
in the sense of circular polarization with increasing frequency in three
non-eclipsing systems: \object{EI~Eri}, \object{DM~UMa} and \object{HD~8358}.
We find clear evidence for coherent plasma emission at 1.4~GHz in the
quiescent spectrum of \object{HD~8358} during the helicity reversal. The
degrees of polarization of the other two systems could also be accounted for
by a coherent emission process. The observations of \object{ER~Vul} revealed
two U-shaped flux spectra at the highest frequencies. The U-shape of the
spectra may be accounted for by an optically thin gyrosynchrotron source for
the low frequency part whereas the high frequency part is dominated by a
thermal emission component.
}

\maketitle

\section{Introduction} \label{intro}

RS~CVn systems are close, chromospherically active binary systems whose
enhanced emission can be detected over a wide range of the spectral domain,
from the X-ray to the radio region. Most of the phenomena observed in these
systems at different wavelengths are attributed to the presence of magnetic
fields generated by a dynamo mechanism.

In general, the radio emission from RS~CVn binary systems is quite variable,
with luminosity levels in the range 10$^{14}$$-$10$^{19}$
erg~s$^{-1}$~Hz$^{-1}$ at centimeter wavelengths (Morris \& Mutel
\cite{Morris88}, Drake et~al. \cite{Drake89}, \cite{Drake92}). Typical
features of the radio emission from these systems are: a) low-level
(quiescent) emission with moderately circularly polarized emission (degree of
circular polarization $\pi_{\rm c} \lesssim$30--40\%) and flat or negative
spectrum (spectral index $\alpha \lesssim 0$, where $S_{\nu} \propto
\nu^{\alpha}$); or b) high-level (flaring) emission, unpolarized or weakly
polarized ($\pi_{\rm c} \lesssim 10$\%), and with positive spectral index
$\alpha$; or c) high-intensity, short-duration outbursts with high degrees of
circular polarization (see Gunn \cite{Gunn96} for a review on RS~CVn binary
systems). The mechanisms generally invoked to account for the observed radio
emission are gyrosynchrotron, synchrotron and coherent (electron-cyclotron
maser or plasma radiation) processes. For reviews on these radio emission
mechanisms see, for instance, Dulk (\cite{Dulk85}) and G\"{u}del
(\cite{Gudel02}).

Gibson (\cite{Gibson84}) suggested that stars that exhibit substantial
circular polarization in their incoherent emission are those with lower
orbital inclination angles, whereas stars with unpolarized or weakly polarized
emission have higher inclination angles. Mutel et~al. (\cite{Mutel87}) studied
a small sample of late-type binaries, mainly of the RS~CVn class, and found
that non-eclipsing (low inclination angle) systems have an average circular
polarization in their quiescent emission significantly larger than that for
eclipsing systems.

The non-thermal nature of the electron distribution responsible for the
flaring emission from RS~CVn systems seems to be well established. The
quiescent emission has been interpreted in two different ways, based on
different assumptions for the distribution of the population of electrons
responsible for the emission observed, namely gyrosynchrotron emission from a
Maxwellian (thermal) distribution or from a power-law (non-thermal) 
distribution (Drake et~al. \cite{Drake89}, \cite{Drake92}; Chiuderi-Drago \&
Franciosini \cite{Chiuderi93}). However, a thermal gyrosynchrotron emission
model with uniform magnetic field predicts a spectral index $\alpha = -8$
after the peak, whereas the spectra observed are much flatter. Thermal
emission in a magnetic field $B$ decreasing as $B \propto r^{-1}$, where $r$
is the distance from the active star, may reproduce the observed quiescent
spectrum, but the magnetic field structure seems unrealistic. Emission from a
non-thermal distribution appears to be the most plausible explanation for the
observed properties of the quiescent emission.

Previous multifrequency polarization observations of several RS~CVn systems
(e.g., Mutel et~al. \cite{Mutel87}, Massi \& Chiuderi-Drago \cite{Massi92}, 
Umana et~al. \cite{Umana93}, White \& Franciosini \cite{White95}) revealed 
properties of both low-level and high-level emission sources, showing a 
reversal in the sense of polarization between 1.4~GHz and 5~GHz for
non-eclipsing systems. In addition, White \& Franciosini (\cite{White95})
found an increase of the degree of polarization with increasing frequency at
high frequencies, at least up to 15~GHz, independently of the shape of the
spectrum. However, the polarization properties of the quiescent emission are
not well reproduced by gyrosynchrotron models. These models predict the
helicity inversion at frequencies where the spectral index is positive,
whereas the observations show this occurrence with a flat or decreasing
spectrum. White \& Franciosini (\cite{White95}) proposed that weak, highly
polarized, coherent plasma emission may be associated with the polarization
inversion observed at low frequencies. Gyrosynchrotron models also predict a
decrease of the degree of circular polarization with increasing frequency at
high frequencies, which is the opposite to the observed tendency.

Thus, some details in the interpretation of the radio emission in active
close binary systems remain still unclear or need additional observational
support. Only a small fraction of RS~CVn systems have measurements of their
circular polarization emission at several frequencies. Multiband centimeter
observations of low-level sources are, hence, important for a correct
interpretation of the radio emission from these systems.

In the present paper we report four-frequency radio observations of a sample 
of eight RS~CVn systems performed at several epochs. The systems are 
\object{DM~UMa}, \object{AY~Cet}, \object{HD~8358}, \object{HD~8357}, 
\object{EI~Eri}, \object{RS~CVn}, \object{$\sigma^{2}$ CrB} and
\object{ER~Vul}, for which no four-frequency radio observations have been
reported previously. The sample selected is representative of the RS~CVn-type
systems in the sense that it contains both eclipsing and non-eclipsing
systems, with different degrees of orbital inclinations, as well as both short
and long-period binaries. The short-period systems contain components that
rotate synchronously whereas, in contrast, no such a synchronization occurs
for the long-period systems.

The paper is organized as follows: Sect.~\ref{sec1} describes the
observations, Sect.~\ref{sec2} presents the results of these observations and
discusses the binary systems individually, and finally Sect.~\ref{sec3}
summarizes the main conclusions.

\section{Observations}\label{sec1} 

We carried out radio observations of the eight RS~CVn systems mentioned above
at 1.4~GHz (L band), 4.9~GHz (C band), 8.5~GHz (X band) and 14.9~GHz (U band).
The observations were made with the Very Large Array (VLA) interferometer of
the NRAO\footnote{The National Radio Astronomy Observatory is a facility of
the National Science Foundation operated under cooperative agreement by
Associated Universities, Inc.} during eight sessions in 1997. The epochs of
observation were May 29 (B/C configuration), June 6 (B/C configuration),
August 24 (C/D configuration), August 31 (C/D configuration), September 27
(C/D configuration), October 2 (C/D configuration), October 3 (C/D
configuration), and October 29 (D configuration). 

The sources were always observed at all frequencies, except in two epochs in
which U band observations could not be performed for some systems due to time 
constraints.  These exceptions were \object{HD~8357} and \object{HD~8358} on
August 24, and \object{DM~UMa} and \object{RS~CVn} on October 3.

The multifrequency observations were performed sequentially. A typical 
observation consisted of 8$-$10 minutes on the target source at each band,
preceded and followed by a phase calibrator observation. We used two
bandwidths of 50 MHz at each band, with central frequencies 1385 and 1465 MHz
at L band, 4835 and 4885 MHz at C band, 8435 and 8485~MHz at X band, and 14915
and 14965~MHz at U band. For flux calibration we used \object{0137+331} and,
on October 3, \object{1331+305}. The assumed flux densities for
\object{0137+331} were 16, 5.5, 3.2 and 1.8~Jy at 1.4, 4.9, 8.5 and 14.9~GHz,
respectively, whereas those for \object{1331+305} were 15, 7.5, 5.2 and 3.4~Jy
at 1.4, 4.9, 8.5 and 14.9~GHz, respectively. For phase calibration the sources
observed were \object{0059+001}, \object{0115$-$014}, \object{0125$-$000},
\object{0149+059}, \object{0423$-$013}, \object{1035+564}, \object{1310+323},
\object{1613+342} and \object{2115+295}. 

The calibration and analysis of the data were carried out using the
Astronomical Image Processing System (AIPS) data reduction software package.
Maps were made with the task IMAGR. We measured the Stokes' parameters I, the
total intensity, and V, the circularly polarized intensity. Positive V
corresponds to right circular polarization, whereas negative V corresponds to
left circular polarization. The degree of circular polarization is defined as
$\pi_{\rm c} \equiv$ V/I. We also used the task DFTPL to analyze the light
curves.

\section{Results and discussion}\label{sec2}

The results of the observations are presented in Table~\ref{table1}. The
measured values of the total intensity, I, and the degree of circular
polarization, $\pi_{\rm c}$, as well as their uncertainties (one-sigma
errors), are listed for each system and epoch of observation. We list the
three-sigma values as upper limits if no radio emission is detected. The eight
binary systems were detected at 4.9~GHz at all epochs of observation, and the
corresponding radio luminosities $L_{\rm 4.9\,GHz}$ at this frequency are
included in the table, with values ranging from 6.1$\times$10$^{14}$ to
3.3$\times$10$^{17}$ erg~s$^{-1}$~Hz$^{-1}$. The last two columns list the
spectral indexes $\alpha$ between 1.4 and 4.9~GHz and between 4.9 and 8.5~GHz.

\begin{table*}[!t]
\caption{Multifrequency observations of the eight RS~CVn systems. All the dates are for year 1997. I represents the total intensity (and its uncertainty) in~mJy, and $\pi_{\rm c}$ the degree of circular polarization (and its uncertainty) in $\%$, for the four frequencies of observation. An asterisk indicates that the total intensity was variable during the corresponding observation at that frequency. Upper limits on I are given as 3$\sigma$ values. The three last columns list the radio luminosity at 4.9~GHz, $L_{\rm 4.9\,GHz}$, and the spectral indexes $\alpha_{\rm 1.4 - 4.9}$ between 1.4 and 4.9~GHz and $\alpha_{\rm 4.9 - 8.5}$ between 4.9 and 8.5~GHz.}
\label{table1}
\begin{center}
\begin{tabular}{llccccccrr}
\hline
\hline
\noalign{\smallskip}
System & Date & I  & \multicolumn{4}{c}{Total Intensity (mJy)} & 
$L_{\rm 4.9\,GHz}$ & \multicolumn{2}{c} {Spectral} \\
       &      & $\pi_{\rm c}$  & \multicolumn{4}{c}{Circular Polarization Degree 
($\%$)} & 
(erg~s$^{-1}$~Hz$^{-1}$) 
& \multicolumn{2}{c} {Index} \\
       &      &    & 1.4 GHz & 4.9 GHz & 8.5 GHz & 14.9 GHz & & 
$\alpha_{\rm 1.4 - 4.9}$ & $\alpha_{\rm 4.9 - 8.5}$ \\
\noalign{\smallskip}
\hline
\noalign{\smallskip}
HD~8357     & May 29& I  & $\leq$ 1.5 & 6.71 $\pm$ 0.04 & 5.92 $\pm$ 0.03 & 3.7 $\pm$ 0.1 & 1.5$\times$10$^{16}$ & & $-$0.2 \\
     &       & $\pi_{\rm c}$  & --- &    3.7 $\pm$ 0.6 & 4.2 $\pm$ 0.5 & 10 $\pm$ 2 & & & \\
\noalign{\smallskip}
     & Aug. 24& I  & ~~~2.4  $\pm$ 0.4$^{(*)}$ & 7.32 $\pm$ 0.04 & 6.27 $\pm$ 0.04 &  ---  & 1.6$\times$10$^{16}$ & 0.9 & $-$0.3  \\
     &       & $\pi_{\rm c}$  & 2.5 $\pm$ 2.7 & 5.1 $\pm$ 0.6 & 11.0 $\pm$ 0.5 & --- & &  & \\
\noalign{\smallskip}
     & Aug. 31& I  & 3.4  $\pm$ 0.5 & 19.42 $\pm$ 0.04 & 25.14 $\pm$ 0.05 & 23.8 $\pm$ 0.2 & 4.2$\times$10$^{16}$ & 1.4 & 0.5 \\
     &       & $\pi_{\rm c}$  & $-$2.6 $\pm$ 2.0  & $-$1.7 $\pm$ 0.2 & $-$1.8 $\pm$ 0.1 & 
$-$0.8 $\pm$ 0.9 & & & \\
\noalign{\smallskip}
     & Sep. 27& I  & $\leq$ 1.5 & 0.28 $\pm$ 0.05 & 0.16 $\pm$ 0.03 & $\leq$ 0.3 & 6.1$\times$10$^{14}$ & & $-$1.0 \\
     &       & $\pi_{\rm c}$  & --- & 20 $\pm$ 10 & 37 $\pm$ 13 & --- & 
& & \\
\noalign{\smallskip}
     & Oct. 29& I  &  0.7  $\pm$ 0.2 & 1.43 $\pm$ 0.04 & 1.48 $\pm$ 0.03 & 0.6  $\pm$ 0.1 & 3.1$\times$10$^{15}$ & 0.6 & 0.06 \\
     &       & $\pi_{\rm c}$  & 4.3 $\pm$ 4.7 & 13 $\pm$ 2 & 17 $\pm$ 2 & 26 $\pm$ 17 &  &  & \\
\noalign{\smallskip}
\hline
\noalign{\smallskip}
EI~Eri     & May 29& I  &  $\leq$ 5.7 &  2.18 $\pm$ 0.06 & ~~~3.42 $\pm$ 0.04$^{(*)}$  &  ~~~0.8  $\pm$ 0.1$^{(*)}$ & 8.3$\times$10$^{15}$  & & 0.8 \\
     &       & $\pi_{\rm c}$  & ---  &  $-$3.2 $\pm$ 1.2 & $-$0.6 $\pm$ 1.1 & $-$23 $\pm$ 13 & &  &  \\
\noalign{\smallskip}
     & Jun. 6 & I  & $\leq$ 4.5 &  4.90 $\pm$ 0.04 &  4.41 $\pm$ 0.04 &  3.1  $\pm$ 0.1 &
1.9$\times$10$^{16}$ & & $-$0.2 \\
     &       & $\pi_{\rm c}$  & --- & 2.2 $\pm$ 0.6 & $-$2.0 $\pm$ 0.8  & $-$5.1 $\pm$ 3.0
& &  & \\
\noalign{\smallskip}
     & Aug. 31& I  &  ~~~8 $\pm$ 2$^{(*)}$ & 22.87 $\pm$ 0.06 & 22.79 $\pm$ 0.05 &  19.5  $\pm$ 0.2 & 8.7$\times$10$^{16}$ & 0.8 & $-$0.01 \\
     &       & $\pi_{\rm c}$  & 43 $\pm$ 13 & 0.26 $\pm$ 0.13 & $-$1.3 $\pm$ 0.2 & $-$1.5 $\pm$ 1.1 &  & & \\
\noalign{\smallskip}
     & Oct. 29& I  & $\leq$ 0.9 & 1.52 $\pm$ 0.05 & 1.31 $\pm$ 0.03 &  0.9  $\pm$ 0.1 & 5.8$\times$10$^{15}$ & & $-$0.3 \\
     &       & $\pi_{\rm c}$  & --- & $-$10 $\pm$ 2 & $-$17 $\pm$ 3 & $-$21 $\pm$ 11  & & & \\
\noalign{\smallskip}
\hline
\noalign{\smallskip}
DM~UMa     & Jun. 6 & I  &  9.2  $\pm$ 0.3 & 14.37 $\pm$ 0.05 & ~~~12.29 $\pm$ 0.08$^{(*)}$ & ~~~2.7
$\pm$ 0.2$^{(*)}$ & 3.3$\times$10$^{17}$ & 0.4 & $-$0.3 \\
     &       & $\pi_{\rm c}$  & 27 $\pm$ 1 &  $-$1.3 $\pm$ 0.2 &  $-$3.9 $\pm$ 0.3  & 
 $-$11 $\pm$ 7 & & & \\
\noalign{\smallskip}
     & Aug. 31& I  & 6.3  $\pm$ 0.7 &   8.09 $\pm$ 0.05&  6.73 $\pm$ 0.06 &   5.2  $\pm$ 0.2 & 1.9$\times$10$^{17}$ & 0.2 & $-$0.3 \\
     &       & $\pi_{\rm c}$  & 5 $\pm$ 1 &  $-$3.0 $\pm$ 0.6 &  $-$5.5 $\pm$ 0.8  & 
 $-$9 $\pm$ 4 & & & \\
\noalign{\smallskip}
     & Sep. 27& I  & $\leq$ 1.2 &  1.4  $\pm$ 0.08 &  1.11 $\pm$ 0.06 &  0.9  $\pm$ 0.2 & 3.2$\times$10$^{16}$ & & $-$0.4 \\
     &       & $\pi_{\rm c}$  & --- & 5 $\pm$ 6 & $-$7 $\pm$ 5 & $-$22 $\pm$ 22 & & 
& \\
\noalign{\smallskip}
     & Oct. 2 & I  & $\leq$ 1.5 & 0.79 $\pm$ 0.05 & 0.62 $\pm$ 0.05 &   0.6  $\pm$ 0.2 & 1.8$\times$10$^{16}$ & & $-$0.4 \\
     &       & $\pi_{\rm c}$  & ---  & $-$6 $\pm$ 4 & $-$37 $\pm$ 11 & $-$43 $\pm$ 33 & & & \\
\noalign{\smallskip}
     & Oct. 3 & I  & $\leq$ 1.5 &  0.82 $\pm$ 0.04  &  0.67 $\pm$ 0.04 &    ---  & 1.9$\times$10$^{16}$ &  & $-$0.4 \\
     &       & $\pi_{\rm c}$  & --- & $-$8 $\pm$ 4  &  $-$15 $\pm$ 4 &  --- & & 
& \\
\noalign{\smallskip}
\hline
\noalign{\smallskip}
$\sigma^{2}$ CrB & Sept. 27& I  & $\leq$ 18 & 1.28 $\pm$ 0.05 & 1.17 $\pm$ 0.04 &  0.5  $\pm$ 0.1 & 7.2$\times$10$^{14}$ & & $-$0.2 \\
      &      & $\pi_{\rm c}$  & --- & $-$9 $\pm$ 3 & $-$15 $\pm$ 3 & 10 $\pm$ 20 & & & \\
\noalign{\smallskip}
     & Oct. 3 & I  & $\leq$ 12  & 4.43 $\pm$ 0.06 & ~~~2.45 $\pm$ 0.05$^{(*)}$ & 0.55 $\pm$ 0.08 & 2.5$\times$10$^{15}$ & & $-$1.1 \\
     &       & $\pi_{\rm c}$  & --- & $-$3.4 $\pm$ 0.9 & $-$3.2 $\pm$ 1.4 & 32 $\pm$ 12 & & & \\
\noalign{\smallskip}
\hline
\noalign{\smallskip}
HD~8358     & May 29& I  & 1.7  $\pm$ 0.5 &  3.09 $\pm$ 0.04  &  2.79 $\pm$ 0.03 & 2.06 $\pm$ 0.09 & 1.6$\times$10$^{16}$ & 0.5 & $-$0.2 \\
     &       & $\pi_{\rm c}$  & 6 $\pm$ 4 & $-$4.2 $\pm$ 1.3  & $-$11 $\pm$ 1 & $-$29  $\pm$ 6 & & & \\
\noalign{\smallskip}
     & Aug. 24& I  & $\leq$ 1.2 & 1.03 $\pm$ 0.04 &   0.88 $\pm$ 0.03 &  ---    & 5.3$\times$10$^{15}$ & & $-$0.3 \\
     &       & $\pi_{\rm c}$  & --- & $-$17 $\pm$ 4 & $-$9.0 $\pm$ 3.4 &   --- & & & \\
\noalign{\smallskip}
     & Aug. 31& I  & ~~~2.2  $\pm$ 0.4$^{(*)}$ &  1.94 $\pm$ 0.04 &  1.69 $\pm$ 0.04 & $\leq$ 0.6 & 1.0$\times$10$^{16}$ & $-$0.1 & $-$0.3 \\
     &       & $\pi_{\rm c}$  & 94 $\pm$ 20 & $-$6 $\pm$ 2 & $-$11 $\pm$ 2 &  --- & & & \\
\noalign{\smallskip}
     & Sep. 27& I  & $\leq$ 2.1 &  1.35 $\pm$ 0.04 & 1.44 $\pm$ 0.03 &  0.8  $\pm$ 0.1 & 7.0$\times$10$^{15}$ & & 0.1 \\
     &       & $\pi_{\rm c}$  & --- & $-$5.9 $\pm$ 2.4  & $-$10 $\pm$ 2 & $-$48 $\pm$ 15 & & & \\
\noalign{\smallskip}
     & Oct. 29& I  & $\leq$ 0.9 &  0.76 $\pm$ 0.03 &  0.79 $\pm$ 0.03 &  $\leq$ 0.3 & 3.9$\times$10$^{15}$ & & 0.07 \\
     &       & $\pi_{\rm c}$  & --- & $-$6 $\pm$ 4 & $-$5.1 $\pm$ 2.7 &  --- & & 
& \\
\noalign{\smallskip}
\hline
\end{tabular}
\end{center}
\end{table*}

\begin{table*}[!t]
 \addtocounter{table}{-1}
 \caption{(continued)}
\begin{center}
\begin{tabular}{llccccccrr}
\hline
\hline
\noalign{\smallskip}
System & Date & I  & \multicolumn{4}{c}{Total Intensity (mJy)} &
$L_{\rm 4.9\,GHz}$ & \multicolumn{2}{c} {Spectral} \\
       &      & $\pi_{\rm c}$  & \multicolumn{4}{c}{Circular Polarization Degree 
($\%$)} &
(erg~s$^{-1}$~Hz$^{-1}$) 
&  \multicolumn{2}{c} {Index}\\
       &      &    & 1.4 GHz & 4.9 GHz & 8.5 GHz & 14.9 GHz & &
$\alpha_{\rm 1.4 - 4.9}$ & $\alpha_{\rm 4.9 - 8.5}$ \\
\noalign{\smallskip}
\hline
\noalign{\smallskip}
ER~Vul      & Aug. 24& I  & $\leq$ 1.5 &  0.59 $\pm$ 0.06 &  0.27 $\pm$ 0.05 & 0.58 $\pm$ 0.07 & 1.8$\times$10$^{15}$ & & $-$1.4 \\
      &       & $\pi_{\rm c}$  & --- & 8 $\pm$ 5 & 15 $\pm$ 11 & 27 $\pm$ 12 & & & \\
\noalign{\smallskip}
      & Aug. 31& I  & $\leq$ 1.2 &  0.52 $\pm$ 0.04 &  0.42 $\pm$ 0.03 & $\leq$ 0.6 & 1.6$\times$10$^{15}$ & & $-$0.4 \\
      &       & $\pi_{\rm c}$  & --- & 8 $\pm$ 6 & 16 $\pm$ 7 & --- & &
& \\
\noalign{\smallskip}
      & Oct. 2 & I  & $\leq$ 6 & 0.64 $\pm$ 0.04  & 0.41 $\pm$ 0.04 &  0.5  $\pm$ 0.1 & 1.9$\times$10$^{15}$ & & $-$0.8 \\
      &       & $\pi_{\rm c}$  & --- & 8 $\pm$ 5 & $-$12 $\pm$ 8 & 12 $\pm$ 20 & & & \\
\noalign{\smallskip}
      & Oct. 29& I  & $\leq$ 3 & 2.32 $\pm$ 0.04 & 3.05 $\pm$ 0.04 & 1.9  $\pm$ 0.1 & 6.9$\times$10$^{15}$ & & 0.5 \\
      &       & $\pi_{\rm c}$  & --- & 2.6 $\pm$ 1.4 & $-$1 $\pm$ 1 & 7 $\pm$ 5 & &
&  \\
\noalign{\smallskip}
\hline
\noalign{\smallskip}
AY~Cet     &May 29& I  & $\leq$ 0.9 &  0.43 $\pm$ 0.04 &  0.18 $\pm$ 0.03 & $\leq$ 0.3 & 3.2$\times$10$^{15}$ & & $-$1.6  \\
     &       & $\pi_{\rm c}$  & --- & $-$13 $\pm$ 8 & $-$30 $\pm$ 17  &  --- & & 
& \\
\noalign{\smallskip}
     & Sep. 27& I  & 1.2  $\pm$ 0.2 &  2.10 $\pm$ 0.05 &  1.84 $\pm$ 0.03 &  1.0  $\pm$ 0.1
& 1.6$\times$10$^{16}$ & 0.4 & $-$0.2 \\
     &       & $\pi_{\rm c}$  & $-$5 $\pm$ 3  & $-$6 $\pm$ 2 & $-$4 $\pm$ 1 & 7 $\pm$ 10  & & &  \\
\noalign{\smallskip}
     & Oct. 2 & I  & $\leq$ 1.2 &   0.78 $\pm$ 0.04 &  0.57 $\pm$ 0.03 & $\leq$ 0.3 &
5.8$\times$10$^{15}$ & & $-$0.6 \\
     &       & $\pi_{\rm c}$  & --- & $-$7 $\pm$ 4 & $-$16 $\pm$ 5 & --- & & & \\
\noalign{\smallskip}
     & Oct. 29& I  & 0.7  $\pm$ 0.2 &  0.29 $\pm$ 0.03  & $\leq$ 0.09 & $\leq$ 0.3 &
2.1$\times$10$^{15}$ & $-$0.7 &  \\
     &       & $\pi_{\rm c}$  & 1 $\pm$ 5 & $-$34 $\pm$ 10 &  --- & --- & & 
& \\
\noalign{\smallskip}
\hline
\noalign{\smallskip}
RS~CVn      & Jun. 6 & I  & $\leq$ 1.5 &  1.60 $\pm$ 0.05 & 1.09 $\pm$ 0.04 & $\leq$ 0.6 &
2.2$\times$10$^{16}$ & & $-$0.7 \\
      &       & $\pi_{\rm c}$  & --- & $-$2.5 $\pm$ 2.0 & $-$8.3 $\pm$ 3.1 & --- &  & & \\
\noalign{\smallskip}
     & Oct. 3  & I  & $\leq$ 1.2 &  0.16 $\pm$ 0.03  & $\leq$ 0.09 & --- 
& 2.2$\times$10$^{15}$ & & \\
      &       & $\pi_{\rm c}$  & --- & 94 $\pm$ 30 & --- & --- & & & \\
\noalign{\smallskip}
\hline
 \end{tabular}
\end{center}
\end{table*}


The emission detected for most of the epochs corresponds to a low-level flux.
Flaring emission at least one order of magnitude above the quiescent level is
detected in three systems, namely \object{HD~8357}, \object{EI~Eri} and
\object{DM~UMa}. We detected circularly polarized radio emission above a
3$\sigma$ level for all the systems, with the only exception of ER~Vul. In
general, the circular polarization measurements reveal weak or moderate levels
of polarization, with values $\mid\pi_{\rm c}\mid \lesssim 30$\% for most of
the detections. However, very highly polarized emission is also present in a
few cases, with $\mid\pi_{\rm c}\mid >$90\%. 

\begin{figure*}
\resizebox{\hsize}{!}{\includegraphics{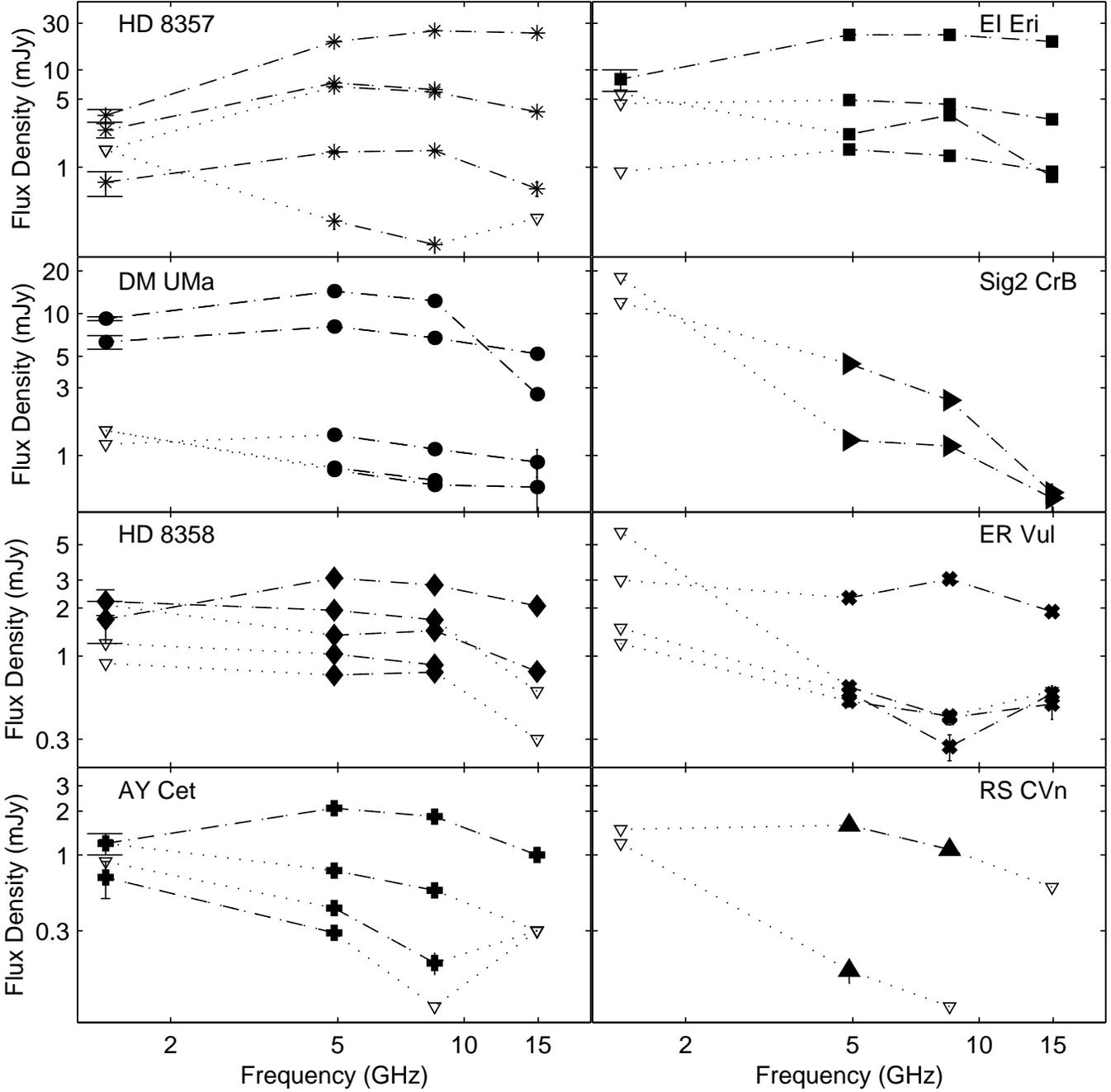}}
\caption{Radio spectra of the eight RS~CVn systems at 1.4~GHz, 4.9~GHz, 8.5~GHz and 14.9~GHz for all the epochs of observation. The axes are logarithmic. The black symbols denote the values detected at different frequencies, and are joined by a dot-dashed line for the corresponding epoch of observation. The white filled, inverted triangles denote the 3$\sigma$ upper limits, which are joined by dotted lines to the values detected for the corresponding epoch of observation. Error bars are also plotted, most of which are smaller than the symbol's size.}
\label{fig2}
\end{figure*}

The radio spectra corresponding to the values listed in Table~\ref{table1} are
plotted in Fig.~\ref{fig2}. We show the flux density as a function of the
frequency of observation for each epoch of observation. The spectral indexes
derived from these values reveal either the quiescent or the flaring nature of
the observed emission. However, we caution that the non-simultaneous nature of
the observations may affect the spectral shapes due to the possible
variability of the emission on short timescales. We analized the radio light
curves of all the observations and found several cases in which variability is
clearly present. The variability is evidenced by the different modulation of
the light curve of a given frequency with respect to the other frequencies at
the same epoch of observation. These cases are indicated with an asterisk
after the frequency at which such variability is found in Table~\ref{table1},
and some of them will be discussed in more detail below.

\begin{figure*}
\resizebox{\hsize}{!}{\includegraphics{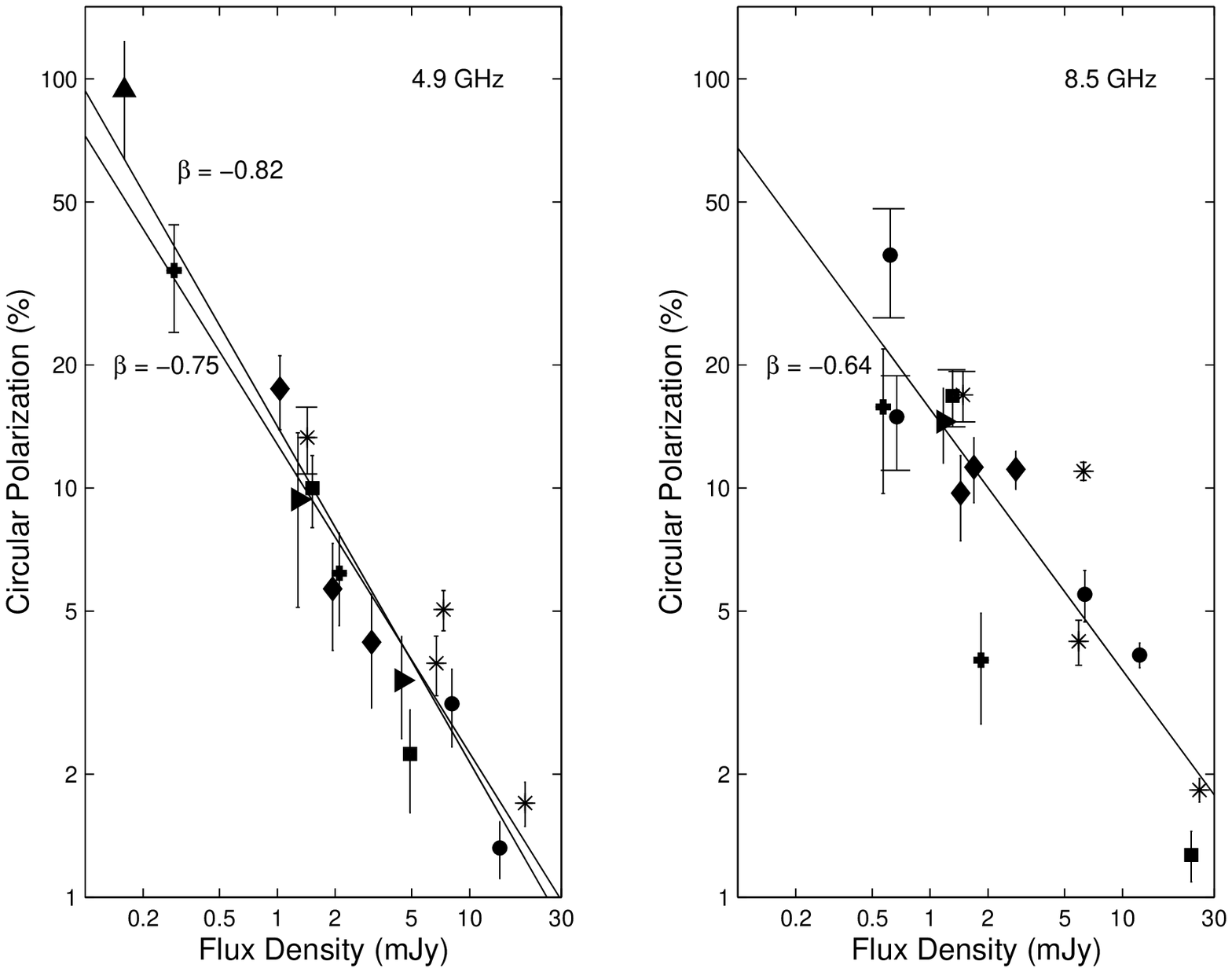}}
\caption{Degree of circular polarization as a function of flux density (in mJy), at 4.9~GHz and 8.5~GHz, for the eight RS~CVn binaries of our sample. The axes are logarithmic. Error bars are plotted for the circular polarization values. Different symbols represent different systems. The symbol used for each binary system is the same as the one used in Fig.~\ref{fig2}. The solid line in the 8.5~GHz observations plot is a fit to the data with a slope $\beta = -0.64$. The solid lines in the 4.9~GHz observations plot are fits to the data with a slope $\beta = -0.82$  when all the systems are included and $\beta = -0.75$ when \object{RS~CVn} is not included. The triangle at the top of the 4.9~GHz plot represents \object{RS~CVn}.}
\label{figpolflux}
\end{figure*}

The circularly polarized emission is mainly detected at 4.9~GHz and 8.5~GHz.
We show the degree of circular polarization, $\pi_{\rm c}$, as a function of
the flux density at these two frequencies in Fig. \ref{figpolflux}. The plots
show a decrease in the fractional circular polarization with increasing level
of radio emission for both frequencies. The anticorrelation between the degree
of circular polarization and the flux density shown in this figure, with a
decrease in $\pi_{\rm c}$ as the flux increases, is in agreement with
gyrosynchrotron emission models in which the quiescent emission is associated
with radiation from an optically thin source, whereas the unpolarized or
weakly polarized flaring emission is associated with a self-absorbed source.
The solid lines in the plots show a least-squares fit to the data using a
function of the form $\pi_{\rm c} \propto$ I$^{\beta}$. The fit to the 8.5~GHz
observations has a slope $\beta$ = $-$0.64, whereas that to the 4.9~GHz
observations has a steeper slope $\beta$ = $-$0.82. However, the circular
polarization of \object{RS~CVn} at 4.9~GHz, unlike the other systems, may be
attributed to a coherent emission mechanism, as we will see later. Thus, we
also fit the same function to the data excluding the circular polarization
measurement of \object{RS~CVn} at 4.9~GHz ($\pi_{\rm c} =$ 94\%, triangle at
the top), and we find a slope $\beta$=$-$0.75, somewhat closer to the slope 
of the fit to the 8.5~GHz observations.

We discuss now the VLA observations of the eight binary systems separately. We
include information on past radio observations of these systems found in the 
literature for the sake of comparison. The main properties of these systems
are listed in Table~\ref{table3}. Data on the photometric, spectroscopic and
orbital properties are taken from Strassmeier et~al. (\cite{Strassmeier93}),
unless a different reference is indicated. Heliocentric distances are from the
astrometric data in the Hipparcos Catalogue (ESA \cite{ESA97}).

\begin{table*}
\caption{Properties of the eight RS~CVn systems.}
\label{table3}
\begin{center}
\begin{tabular}{lcccccccc}
\hline
\hline
System & Spectral Types & Masses$^{a}$ & $P_{\rm orb}^{b}$ & $P_{\rm rot}^{c}$ & 
Eccentricity & $i^{d}$ & Eclipse & Distance \\
 &  & (M$_{\sun}$) & (days) & (days) & & ($^{\circ}$) & & (pc) \\
\hline
HD~8357$^{e}$ & G7 V$^{e}$ / K1 IV$^{e}$ & 0.92 / 1.12 & 14.3023$^{e}$ & 12.245 & 0.185$^{e}$ 
& 30$^{e}$ &  none & 45 \\
EI~Eri & G5 IV / ? & $\geq$1.4 / $\geq$0.53 & 1.947227 & 1.945 & 0.0 & 46 & none 
& 56 \\
DM~UMa & K0$-$1 IV$-$III / ?  & f(m)=0.011 & 7.4949 & 7.478 & 0.02 & $\approx$ 40 
& none & 139  \\
$\sigma^{2}$ CrB & F6 V / G0 V & 1.12 / 1.14 & 1.139791 & 1.1687 & 0.022 & 28  
& none & 22  \\
HD~8358 & G5 / G5 & $\geq$0.9 / $\geq$0.8  & 0.515782 & 0.52006 & 0.0 & $\approx$ 30
  & none & 66 \\
ER~Vul & G0 V / G5 V & 1.10 / 1.05 & 0.698095 & 0.6942 & 0.0 & 67 & partial & 50 \\
AY~Cet & WD / G5 III & 0.55 / 2.09 & 56.824 & 77.22 & 0.0 & 26 & none & 79 \\
RS~CVn & F4 IV / G9 IV  & 1.41 / 1.44  & 4.79785 & 4.7912 & 0.0 & 87 & total 
& 108 \\
\hline
 \end{tabular}
\end{center}
{
$^{a}$ Masses are given in the same order as for spectral types, 
where f(m) indicates that the mass function is given. \\
$^{b}$ Orbital period. \\
$^{c}$ Rotation (photometric) period. \\
$^{d}$ Inclination of the pole of the orbital plane or of
the rotation axis. \\
$^{e}$ Data from Fekel (\cite{Fekel96}). \\
}
\end{table*}


\subsection{HD~8357}

In a microwave survey at X band, Slee et~al. (\cite{Slee87}) observed 
\object{HD~8357} and found a maximum daily average flux density of 11.9~mJy 
and a median daily average flux density of 8.9~mJy. Observations at C band
were conducted by Morris \& Mutel (\cite{Morris88}) and also by Drake et~al.
(\cite{Drake89}), who reported values of 2.91~mJy and 2.35~mJy, respectively.

We observed \object{HD~8357} at five epochs. \object{HD~8357} is the source
with the largest difference in our sample between the extreme values of
$L_{\rm 4.9\,GHz}$, of about two orders of magnitude. Weak circular
polarization was detected at C, X and U bands, whereas no circular
polarization flux was detected above a 3$\sigma$ level of $\mid\pi_{\rm c}\mid
\leq$14\% at L band. We detected the highest levels of emission on August 31,
which correspond to a flaring spectrum with positive spectral indexes
$\alpha_{\rm 1.4 - 4.9}=$ 1.4 and $\alpha_{\rm 4.9 - 8.5}=$ 0.5, and a quite
flat spectrum for frequencies higher than 8.5~GHz. The degrees of circular
polarization at C and X bands are negative during this flare, in contrast with
the positive values of the other epochs. These low negative $\pi_{\rm c}$ of
less than 2\%, together with the positive spectral indexes, are consistent
with gyrosynchrotron emission from an optically thick source.

Our observations of \object{HD~8357} agree with the trend found by White \&
Franciosini (\cite{White95}), that is, $\pi_{\rm c}$ tends to increase with
frequency at high frequencies. In particular, we observed quiescent emission
on May 29, with negative spectral indexes $\alpha_{\rm 4.9 - 8.5} =$ $-$0.2
and $\alpha_{\rm 8.5 - 14.9} =$ $-$0.8, while the degrees of circular
polarization increased with frequency, that is, $\pi_{\rm c} =$ 3.7\%,
$\pi_{\rm c} =$ 4.2\% and $\pi_{\rm c} =$ 10\% at 4.9~GHz, 8.5~GHz and
14.9~GHz, respectively, in conflict with predictions by gyrosynchrotron
models.

\begin{figure}
\resizebox{\hsize}{!}{\includegraphics{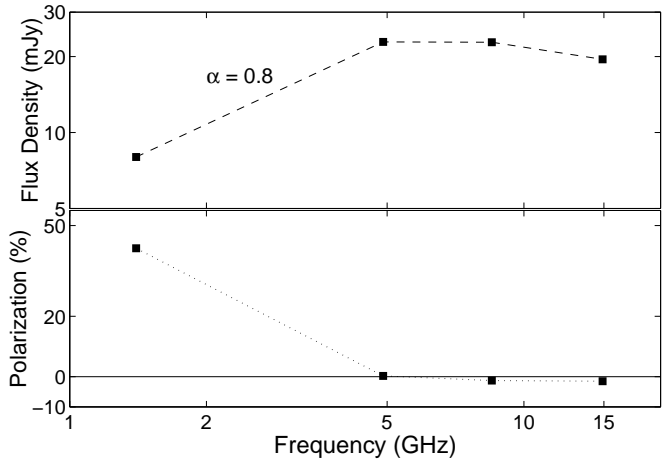}}
\caption{Flux density and degree of circular polarization of \object{EI~Eri}
on August 31, at 1.4~GHz, 4.9~GHz, 8.5~GHz and 14.9~GHz. The plot at the top shows the radio spectrum of \object{EI~Eri} with spectral index $\alpha$ = 0.8 between 1.4 and 4.9~GHz. Both axes are logarithmic. The bottom plot displays the circular polarization (in percent) as a function of the frequency of observation (logarithmic axis). The $\pi_{\rm c}$ values at 4.9~GHz and 14.9~GHz are below a 3$\sigma$ level. There is a reversal in the sense of polarization between 1.4 and 8.5~GHz.}
\label{eiepol}
\end{figure}


\subsection{EI~Eri}

Mutel \& Lestrade (\cite{Mutel85a}) observed \object{EI~Eri} at C band and 
measured a radio flux of 4.3~mJy. Slee et~al. (\cite{Slee88}) found fluxes of
4.7~mJy at C band and 18.6~mJy at X band. Slee et~al. (\cite{Slee87}) reported
a maximum and median daily average fluxes of 21.1~mJy and 11.4~mJy,
respectively, in their survey at X band. VLBI observations at S band (2.3~GHz)
were performed by White et~al. (\cite{White90}), who found a flux density of
15~mJy. The decay phase of a flare was observed by Fox et~al. (\cite{Fox94})
at X, C and L bands, with maximum fluxes of $\sim$20~mJy ($\pi_{\rm c} =$
+5\%), 14~mJy ($\pi_{\rm c} =$ +20\%), and 8.8~mJy ($\pi_{\rm c} =$ +51\%),
respectively.

We observed \object{EI~Eri} at four epochs. This system was always detected at
C, X and U bands, with low polarization fluxes. At L band there was only one
detection, with a relatively high $\pi_{\rm c}$. On October 29 there is an
increase of $\pi_{\rm c}$ with increasing frequency. The highest radio
luminosities correpond to the observations on August 31, in which we detected
a flare with a reversal in the sense of circular polarization between 1.4 and
8.5~GHz. The flux and polarization spectra of the August 31 observations are
plotted in Fig.~\ref{eiepol}. The flux spectrum (plot at the top) shows a
positive spectral index $\alpha_{\rm 1.4 - 4.9} =$ 0.8.

\begin{figure}
\resizebox{\hsize}{!}{\includegraphics{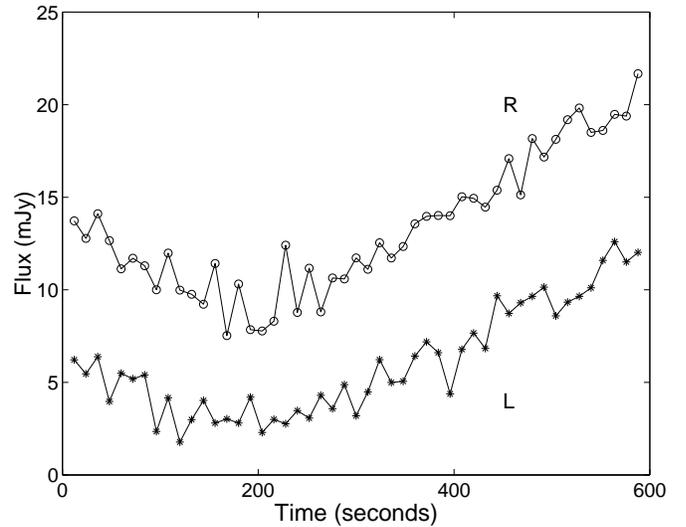}}
\caption{Flux as a function of time for the right circular polarization R (circle symbols) and the left L circular polarization (asterisk symbols) at 1.4~GHz of \object{EI~Eri} on August 31. The flux variations are plotted in 12 second intervals.}
\label{eie31}
\end{figure}

The inversion in the sense of circular polarization could be explained in
terms of a gyrosynchrotron model in which the reversal takes place when the
frequency of observation passes from a low frequency optically thick source 
to a high frequency optically thin source. The V fluxes for gyrosynchrotron
emission can be either positive or negative for each mode of propagation, the
x-mode (in the optically thin part) and the o-mode (in the optically thick
part), depending on the magnetic field orientation with respect to the
observer's line of sight. However, typical values of the degree of
polarization for flaring, incoherent gyrosynchrotron emission are
$\mid\pi_{\rm c}\mid \lesssim$ 10\%, whereas the value measured at L band is
relatively high, $\pi_{\rm c} =$ 43\%. Thus, we also consider that a coherent
emission process, such as electron-cyclotron maser or plasma emission, may be
present at this band.

White \& Franciosini (\cite{White95}) studied the reversal in the sense of
polarization between low and high frequencies, and interpreted the difference 
between the flux evolution of the left and right polarizations at low
frequencies as owing to the presence of plasma emission, with rapid
variability of the coherent component on timescales of $\sim$10 seconds. In
order to investigate if such a emission process is also present in our
observations, we plot the time evolution of the right circular polarization
(R) and the left circular polarization (L) fluxes at 1.4~GHz in 12 second
intervals in Fig.~\ref{eie31}. The overall evolution of the R and L components
is practically the same, displaying a decreasing intensity trend in both
profiles during the first third of the observing time interval, and increasing
afterwards. The close resemblance between the evolution of the R and L
components is not expected if only one of the polarization components is due
to a coherent emission process, in which case the modulation of one component 
would be different than the other. However, such a high $\pi_{\rm c}$ in the
optically thick part of the spectrum can only be accounted for by coherent
emission.

\begin{figure}
\resizebox{\hsize}{!}{\includegraphics{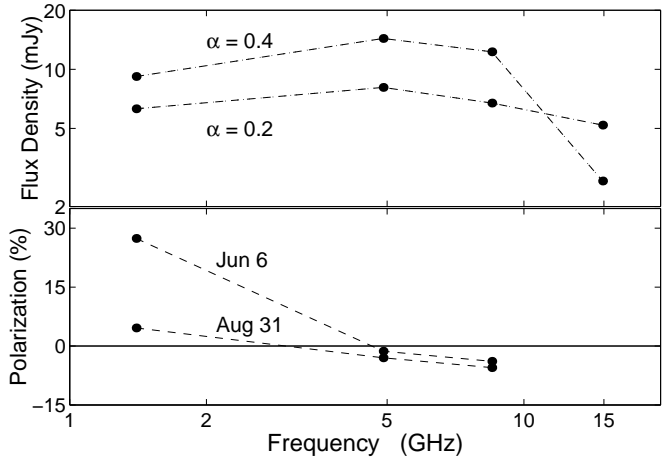}}
\caption{Flux density and degree of circular polarization of \object{DM~UMa} for two epochs, June 6 and August 31, at 1.4~GHz, 4.9~GHz, 8.5~GHz and 14.9~GHz. The plot at the top shows the radio spectra of \object{DM~UMa}, with spectral indexes $\alpha$ = 0.35 (June 6) and $\alpha$ = 0.2 (August 31) between 1.4 and 4.9~GHz. Both axes are logarithmic. The bottom plot displays the circular polarization (in percent) as a function of the frequency of observation (logarithmic axis). There is a reversal in the sense of polarization between 1.4 and 4.9~GHz for both epochs.}
\label{dmpol}
\end{figure}


\subsection{DM~UMa}

\object{DM~UMa} is the most distant system in our sample. The radio emission
of \object{DM~UMa} has been measured at C band by Mutel \& Lestrade
(\cite{Mutel85a}), who observed a flux of 3.0~mJy, and by Gunn et~al.
(\cite{Gunn94}), who reported a value of 2.04~mJy.

We observed \object{DM~UMa} at five epochs, in which the fluxes detected at C
and X bands were left-handed circularly polarized, whereas its flux at U band
was always unpolarized and the two detections at L band were right-handed
circularly polarized. The highest levels of emission correspond to the
observations performed on June 6 and August 31, for which emission at all four
frequencies was detected. The corresponding radio spectra for these two
epochs, shown in Fig. \ref{dmpol}, are typical of moderate flares, with maxima
between 1.4 and 8.5~GHz. \object{DM~UMa} is the source with the largest radio
luminosity in our sample, $L_{\rm 4.9\,GHz} =$ 3.3$\times$10$^{17}$
erg~s$^{-1}$~Hz$^{-1}$ for the June 6 observation. In contrast, the
observations at the other three epochs are characterized by a lower flux
level, $L_{\rm 4.9\,GHz}$ one order of magnitude smaller, with negative
spectral indexes typical of quiescent states.

\begin{figure}
\resizebox{\hsize}{!}{\includegraphics{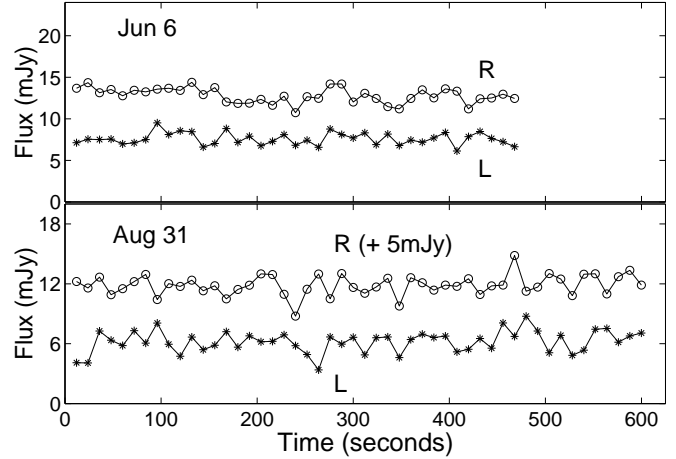}}
\caption{Flux as a function of time at 1.4~GHz for the right circular polarization R (circle symbols) and the left circular polarization L (asterisk symbols) of \object{DM~UMa} on June 6 and August 31. The flux evolution is plotted in 12 second intervals. Polarization R in the bottom plot has been arbitrarily increased by 5~mJy for the sake of comparison with polarization L.}
\label{dmtot}
\end{figure}

We found a helicity reversal between 1.4~GHz and 4.9~GHz for the June 6 and
August 31 observations. The flux spectra and the degree of circular
polarization as a function of frequency are plotted in Fig.~\ref{dmpol}. The
positive spectral index between these two frequencies, $\alpha_{\rm 1.4 -
4.9}$ = 0.4 on June 6 and $\alpha_{\rm 1.4 - 4.9}$ = 0.2 on August 31,
evidences that the source is not quiescent during the helicity reversal. The
degree of circular polarization increases with frequency after the helicity
reversal for both epochs. The three-sigma upper limits on $\pi_{\rm c}$ at
14.9~GHz, 22\% and 12\% on June 6 and August 31, respectively, do not exclude
that the increase continues at the highest frequency of observation. On August
31, $\pi_{\rm c}$ is small, less than 6\%, at L, C and X bands, and this
suggests that the inversion in the sense of polarization is consistent with a
gyrosynchrotron process in which the sense of polarization observed is due to
an optically thick source at low frequencies and an optically thin source at
high frequencies. On June 6, the degree of polarization is also small at C and
X bands, less than 4\%, but in contrast $\pi_{\rm c} \simeq$ 27\% at L band.
Such a $\pi_{\rm c}$ seems to be too high to be produced by gyrosynchrotron
emission from an optically thick source, so a coherent emission mechanism
should be invoked. Analogously to \object{EI~Eri}, we plot the time evolution
of the right circular polarization (R) and the left circular polarization (L)
of \object{DM~UMa} at 1.4~GHz in Fig.~\ref{dmtot}, for the June 6 and August
31 observations. We find no evidence for a different modulation in the flux
evolution of the two components of polarization, as for \object{EI~Eri} above.

\subsection{$\sigma^{2}$ CrB}

This binary, the nearest in our sample, was first detected as radio source by
Spangler et~al. (\cite{Spangler77}), at C band, with a mean flux of 11~mJy.
\object{$\sigma^{2}$ CrB} was also detected at C band by Florkowski et~al.
(\cite{Florkowski85}), who reported two measurements of 8.4~mJy and 19.6~mJy,
by Morris \& Mutel (\cite{Morris88}), with a flux of 8.5~mJy, by Mutel et~al.
(\cite{Mutel85b}), with a flux of 14$-$16~mJy, by L\'{e}fevre et~al.
(\cite{Lefevre94}), with a mean flux density from 8~mJy to 31~mJy, and by Gunn
et~al. (\cite{Gunn94}), who measured a flux of 6.69~mJy. At S band, Turner
(\cite{Turner85}) measured a value of 6~mJy. At L band, the flux detected by
Kuijpers \& van der Hulst (\cite{Kuijpers85}) and by van den Oord et~al.
(\cite{Oord88}) was 9.3~mJy and 8.2~mJy, respectively. At X band, Paredes
et~al. (\cite{Paredes87}) observed a flux of 32~mJy, and Estalella et~al.
(\cite{Estalella93}) fluxes of 14~mJy and 19~mJy at two separate epochs. VLBI
observations by Lestrade et~al. (\cite{Lestrade92}) measured a flux of 12~mJy
at X band, as well as flux levels in the range 7$-$50~mJy at C band.
Observations of the radio emission at L, C and X band, coordinated with X-ray
and UV observations, were performed by Stern et~al. (\cite{Stern92}) and Osten
et~al. (\cite{Osten00}, \cite{Osten03}). Stern et~al. (\cite{Stern92}) found
mean quiescent fluxes of 2.2~mJy at X band, 4.6~mJy at C band, and 9.3~mJy at
L band, as well as strong flaring emission about ten times higher than the
quiescent emission near the end of the observation at the three bands. They
found no circular polarization greater than the 2--3$\sigma$ level. Osten
et~al. (\cite{Osten00}) observed large variations in the radio emission of
\object{$\sigma^{2}$ CrB}, with flux peaks of $\sim$30~mJy ($\pi_{\rm c}
\simeq$ +30\%) at L band, $\sim$81~mJy ($\pi_{\rm c} \lesssim$ +15\%) at C
band, and $\sim$110~mJy (practically unpolarized) at X band. Finally, Osten
et~al. (\cite{Osten03}) reported fluxes of some few mJy at C and X bands, with
negative $\pi_{\rm c}$.

We detected low flux levels in our two-epoch observations of 
\object{$\sigma^{2}$ CrB}, with radio luminosities in the range $L_{\rm
4.9\,GHz}$ $\sim$10$^{14}$$-$10$^{15}$ erg~s$^{-1}$~Hz$^{-1}$, among the
lowest in our sample. The circular polarization was always negative, with
values typical of quiescent emission. On September 27 there is an increase of
the degree of circular polarization with increasing frequency, contrary to
what gyrosynchrotron models predict. The radio spectra for the two epochs of
observation are consistent with an optically thin emitting source. The upper
limits on the flux at L band are also compatible with this interpretation.

\begin{figure}
\resizebox{\hsize}{!}{\includegraphics{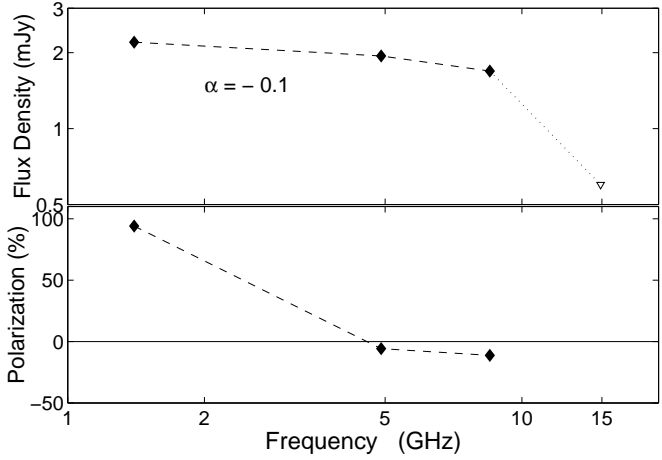}}
\caption{Flux density and degree of circular polarization of \object{HD~8358} on August 31, at 1.4~GHz, 4.9~GHz, 8.5~GHz and 14.9~GHz. The plot at the top shows the radio spectrum of \object{HD~8358} with spectra index $\alpha$ = $-$0.1 between 1.4 and 4.9~GHz. Both axes are logarithmic. The bottom plot displays the circular polarization (in percent) as a function of the frequency of observation (logarithmic axis). There is a reversal in the sense of polarization between 1.4 and 4.9~GHz.}
\label{hd58pol}
\end{figure}


\subsection{HD~8358}

This binary system is a fast rotator, with the shorter rotation period in our
sample (0.52006 days). \object{HD~8358} exhibited low levels of emission at C
band in the past. Drake et~al. (\cite{Drake86}, \cite{Drake92}) reported a
flux of 2.72~mJy with a left-hand circular polarization of $-$9.2\%, and
Morris \& Mutel (\cite{Morris88}) a flux of 3.23~mJy.

\begin{figure}
\resizebox{\hsize}{!}{\includegraphics{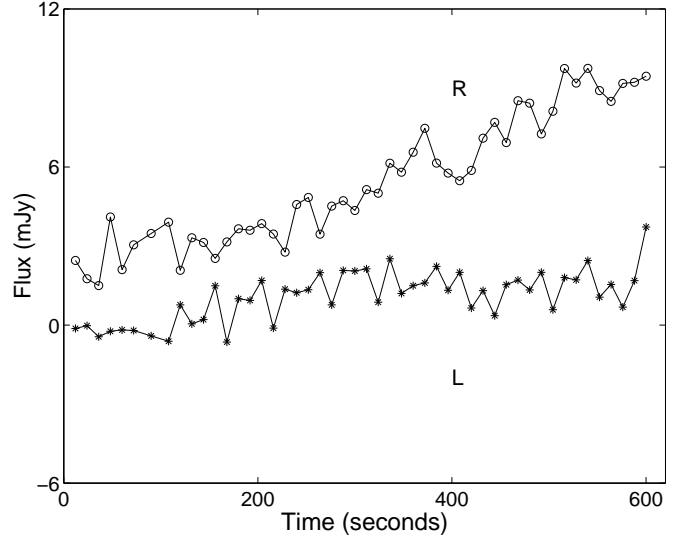}}
\caption{Flux as a function of time for the right circular polarization R (circle symbols) and the left L circular polarization (asterisk symbols) at 1.4~GHz of \object{HD~8358} on August 31. The flux variations are plotted in 12 second intervals.}
\label{h5831}
\end{figure}

We observed \object{HD~8358} at five epochs. The flux measured was always low,
below 3.1~mJy. At C, X and L bands we detected negative circular polarization,
within the range 4\% $\leq \mid \pi_{\rm c} \mid \leq$ 29\%. The degrees of
circular polarization at C band, $-$4\%, $-$6\% and $-$17\%, indicate weakly
polarized emission at this band, in agreement with the measurement by Drake
et~al. (\cite{Drake92}) above. On May 29, the flux spectrum for frequencies
larger than 4.9~GHz corresponds to an optically thin source with spectral
indexes $\alpha_{\rm 4.9 - 8.5} =$ $-$0.2 and $\alpha_{\rm 8.5 - 14.9} =$
$-$0.5, whereas the degree of circular polarization increases with frequency,
$\pi_{\rm c} =$ $-$4\% at 4.9~GHz, $\pi_{\rm c} =$ $-$11\% at 8.5~GHz and
$\pi_{\rm c} =$ $-$29\% at 14.9~GHz.

The source displayed a helicity reversal between L and C bands on August 31,
with a highly polarized, right-hand emission, $\pi_{\rm c} =$ 94\%, at L band,
the only detection of polarization at this band in our observations. We show
this reversal in Fig. \ref{hd58pol}. The spectrum at the top of the figure is
compatible with quiescent emission from an optically thin source at all the
frequencies of observation, with a negative spectral index $\alpha_{\rm 1.4 -
4.9}$ = $-$0.1. The brightness temperature of the source is given by $T_{\rm
b} = 1.77\times10^{9} (S_{\nu}/{\rm mJy})(\nu/{\rm GHz})^{-2}(\theta/{\rm
mas})^{-2}$~K, where $\theta$ is the angular size in milliarcseconds. At $\nu
=$ 1.4~GHz, and assuming a radius $R$ for the emitting region, we find $T_{\rm
b} = 10^{11} (R/$R$_{\sun})^{-2}$~K. The high degree of polarization and the
high brightness temperature at L band may be accounted for by a coherent
emission process. 

The time evolution of the R and L circular polarization flux at 1.4~GHz on
August 31 is plotted in Fig.~\ref{h5831}. There is a different evolution of
the polarization components, with a rise trend in the R component for most of
the observing time, whereas the evolution of the L component is quite flatter.
The contrast is more evident in the second half of the observation. The
different modulation and the high degree of polarization strongly suggests
that coherent emission is present at this frequency. In particular, we
consider plasma emission. The scenario proposed by White \& Franciosini
(\cite{White95}) to explain the helicity reversal, that is, that the observed
reversal at low frequencies is due to the presence of two components, a
gyrosynchrotron emitting source that is weakly polarized or unpolarized at
1.4~GHz, and a weak but highly polarized coherent component (plasma emission)
that is responsible for the high polarization observed, is likely to apply to
the emission of \object{HD~8358} at 1.4~GHz, as evidenced by the different
evolution of the polarization components. Thus, coherent emission at 1.4~GHz
in the form of plasma emission (or alternatively electron-cyclotron maser
emission) is the most likely explanation for the observed helicity reversal.


\subsection{ER~Vul}

Observations at C band of this eclipsing binary were performed by Drake et~al.
(\cite{Drake86}) and by Morris \& Mutel (\cite{Morris88}), who found values of
4.97~mJy and 2.7~mJy, respectively. Rucinski (\cite{Rucinski92}) observed
\object{ER~Vul} in 1990 and 1991, and listed values between 0.66$-$1.9~mJy at
L band and 2.82$-$7.23~mJy at C band for the 1990's observations, and
0.23$-$3.23~mJy at X band and values of 0.65~mJy and 1.61~mJy at C band for
the 1991's observations. This author found no appreciable circular
polarization at any time. At X band, Rucinski (\cite{Rucinski98}) measured
fluxes below 3.5~mJy during his 1995's observations. At L band, flux levels
below $\sim$2.2~mJy were observed by Osten et~al. (\cite{Osten02}), with an
average $\pi_{\rm c} = -8$\%$\pm19$\%.

We observed \object{ER~Vul} during four epochs in which its radio emission was
detected at C and X bands and, except on August 31, also at U band, while the
source was never detected at L band. The radio spectrum of \object{ER~Vul} on
October 29 shows a flaring emission, with a positive spectral index
$\alpha_{\rm 4.9 - 8.5} =$ 0.5. In contrast, lower, quiescent emission levels
were observed at the other three epochs, with negative $\alpha_{\rm 4.9 -
8.5}$ and luminosities ranging $L_{\rm 4.9\,GHz} = 1.6-1.9\times10^{15}$
erg~s$^{-1}$~Hz$^{-1}$. \object{ER~Vul} is the only source in our sample for
which the circular polarization could not be detected above a 3$\sigma$ level
at any of the frequencies of observation. This is not unexpected for such a
high orbital inclination system if its emission is accounted for by a
gyrosynchrotron emission process (Mutel et~al. \cite{Mutel87}). However, the
source was very weak for most of the observations, with fluxes below 1~mJy,
which might also explain the non detection of polarization.

On August 24 and October 2, the quiescent emission of \object{ER~Vul} is
characterized by a U-shaped spectrum at the highest frequencies of observation
(see Fig.~\ref{fig2}), suggesting two spectral components. The slope is
negative up to 8.5~GHz and then becomes positive, with values $\alpha_{\rm 4.9
- 8.5} =$ $-$1.5 and $\alpha_{\rm 8.5 - 14.9} =$ 1.3 for August 24, and
$\alpha_{\rm 4.9 - 8.5} =$ $-$0.8 and $\alpha_{\rm 8.5 - 14.9} =$ 0.4 for
October 2. The negative slope and the low flux level are compatible with
gyrosynchrotron emission from an optically thin source, whereas a different
emission mechanism should be responsible for the highest frequency part of the
spectrum. We find no evidence for a different modulation of the radio light
curve at 14.9~GHz with respect to the light curves at the other frequencies.
Such a difference would account for the U-shaped feature as due to variability
at the highest frequency of observation. However, only simultaneous
multifrequency observations may ultimately confirm the U-shaped spectra
observed.

Similar U-shaped spectra have been observed in the quiescent emission of some
radio active dMe stars like \object{AU Mic} (Cox \& Gibson, \cite{Cox85}), 
the binary \object{UV~Ceti} and other dMe stars (G\"{u}del \& Benz,
\cite{Gudel96}). The low-frequency part of the U-shaped spectra of these
objects was interpreted as optically thin gyrosynchrotron emission, whereas
the high-frequency component was attributed to thermal gyroresonance emission
from the X-ray emitting plasma (G\"{u}del \& Benz \cite{Gudel89}, Cox \&
Gibson \cite{Cox85}). We thus investigate if the high-frequency component of
the spectra of \object{ER~Vul} may also be due to thermal emission.

\object{ER~Vul} has been detected in the X-ray band, with luminosities
$L_{\rm X} =$ 6$-$9$\times$10$^{30}$ erg~s$^{-1}$ (White et~al.
\cite{White87}), and $L_{\rm X} =$ 3.76$\times$10$^{30}$ erg~s$^{-1}$ 
(Dempsey et~al. \cite{Dempsey93}), both assuming a distance $d =$ 45 pc, and
$L_{\rm X} =$ 5.0$\times$10$^{30}$ erg~s$^{-1}$ (Osten et~al. \cite{Osten02}),
which is the measurement reported by Dempsey et~al. (\cite{Dempsey93}) but
assuming the Hipparcos distance. White et~al. (\cite{White87}) used a
two-component thermal plasma model to fit the spectrum, with temperatures of
6$\times$10$^{6}$ K and 4$\times$10$^{7}$ K. Gunn \& Doyle (\cite{Gunn97})
derived a large filling factor of 0.53, as well as a large chromospheric
thickness of $\leq$ 0.42 R$_{\sun}$, for the secondary component of
\object{ER~Vul}. Despite the high orbital inclination of \object{ER~Vul}, no
eclipses have been observed so far in the UV, radio and X-ray domain (Osten
et~al. \cite{Osten02}). The lack of either orbital or rotational modulation in
the X-ray band seems to indicate that the emission likely originates in a
region larger than the stellar size. Both components of \object{ER~Vul} have a
radius $R_{\star} =$ 1.07 R$_{\sun}$, and are separated by a distance $\sim$2
R$_{\sun}$. Thus, assuming that the X-ray emitting plasma region encompasses
both stars, the resulting radius of the region is $R \simeq$ 3 R$_{\sun}$. We
consider that the total flux observed at $\nu =$ 14.9~GHz is due to the
contribution of two emitting sources, thermal plus non-thermal. The maximum
flux emitted by the non-thermal, optically thin gyrosynchrotron source can be
roughly estimated by assuming $\alpha_{\rm 4.9 - 8.5}$ = $\alpha_{\rm 8.5 -
14.9}$, where $\alpha_{\rm 4.9 - 8.5} = -1.4$. This gives a flux from the
non-thermal source of 0.12~mJy on August 24 and 0.26~mJy on October 2. Hence,
the thermal emission fluxes would be 0.46~mJy on August 24 and 0.24~mJy on
October 2, resulting in lower limits on the brightness temperatures of $T_{\rm
b} =$ 1.2$\times$10$^{7}$ K and $T_{\rm b} =$ 6.2$\times$10$^{6}$ K,
respectively. On the other hand, assuming that the flux observed at $\nu =$
14.9~GHz is due entirely to the thermal source, we obtain upper limits 
$T_{\rm b} =$ 1.5$\times$10$^{7}$ K on August 24 and $T_{\rm b} =$
1.3$\times$10$^{7}$ K on October 2. The range of brightness temperatures
obtained are consistent with a thermal emitting source. Even decreasing the
adopted size of the emitting region by a factor $\sim$2, which would result in
a size similar to that of the secondary component plus chromosphere assuming
maximum thickness, the brightness temperatures would be compatible with
thermal emission as well.

On the other hand, the X-ray emission from \object{ER~Vul} might also be
located at one of the poles of the star, and this could also explain the lack
of modulation in the X-ray light curve. An example of X-ray emitting polar
source is \object{AB~Dor} (Maggio et~al. \cite{Maggio00}). \object{Algol}'s
X-ray and radio emission has also been modeled in terms of a corona confined
in the polar regions, with a compact flaring component smaller than the star
and a more extended quiescent corona as large as, or larger than, the star
(Favata et~al. \cite{Favata00}, Mutel et~al. \cite{Mutel98}). VLBI images of
\object{UV~Ceti} (Benz et~al. \cite{Benz98}) reveal that this binary is
another example of radio emitting polar source. Therefore, it would be
possible to obtain brightness temperatures for \object{ER~Vul} consistent with
thermal emission for sizes of the emitting X-ray source similar to those
discussed above.

On August 31, the flux levels at C and X bands are similar to those for the
two epochs above, but at U band we could only measure an upper limit of
0.6~mJy. Thus, we can not confirm a U-shaped flux spectrum for this epoch as
well, though the upper limit on the flux does not rule out this possibility.


\subsection{AY~Cet}

This single-lined binary was observed at C band by Mutel \& Lestrade
(\cite{Mutel85a}), who reported fluxes of 1.1~mJy and 5.2~mJy, and by Morris
\& Mutel (\cite{Morris88}), who measured a value of 0.7~mJy. Simon et~al.
(\cite{Simon85}) observed \object{AY~Cet} at L, C and U bands, and measured
highly variable and highly polarized emission at L band, with fluxes ranging
between 10.6 and 17.7~mJy and negative degrees of polarization 62\% $\leq$
$\mid\pi_{\rm c}\mid$ $\leq$ 86\%, whereas at C band the flux was 2.5~mJy with
$\pi_{\rm c} =$ +5\%, and at U band they found an upper limit of $\sim$5~mJy.
These authors also measured an unpolarized flux of 18.2~mJy at C band two days
before their multifrequency observations.

Our four-epoch observations of \object{AY~Cet} are characterized by relatively
low fluxes at all bands. We detected left-hand circular polarization at C and
X bands, 4\% $\leq$ $\mid$ $\pi_{\rm c}\mid$ $\leq$ 34\%. The flux spectra and
the values of $\pi_{\rm c}$ are compatible with gyrosynchrotron radiation. We
detected \object{AY~Cet} at the four bands on September 27, with a degree of
polarization $\pi_{\rm c} =$ $-$6\% at C band for a flux of 2.1~mJy. This
helicity is the opposite to that found by Simon et~al. (\cite{Simon85}),
$\pi_{\rm c} =$ +5\%, for a similar flux (2.5~mJy). Mutel et~al.
(\cite{Mutel87}) found that for non-eclipsing systems the helicity at a given
frequency is nearly the same over many years. Such a correlation is expected
for \object{AY~Cet} since it is a low-inclination (non-eclipsing) system.
However, the spectrum's slope between L and C bands is $\alpha \simeq$ 0.4 for
our observations, whereas that of Simon et~al.'s observations is clearly
negative, $\alpha \approx$ $-$1.4. As pointed out by these authors, a coherent
mechanism was likely acting during their L band observations. In contrast, a
gyrosynchrotron emission mechanism is consistent with our observations, which
explains the difference in helicity for the same frequency.


\subsection{RS~CVn}

The eclipsing binary \object{RS~CVn} has the highest orbital inclination in
our sample ($i =$ 87$^{\circ}$). \object{RS~CVn} was observed at C band by
Gibson \& Newell (\cite{Gibson79}), who found a flux of 7.2~mJy, by Florkowski
et~al. (\cite{Florkowski85}), who reported a value of 2.96~mJy, and by Gunn
et~al. (\cite{Gunn94}), with a flux of 1.86~mJy.


We observed \object{RS~CVn} at two epochs. The source was detected at X and C
bands on June 6, with a negative spectral index $\alpha_{\rm 4.9 - 8.5} =
-0.7$. No circular polarization was found at these bands above a 3$\sigma$
level. Unpolarized or weakly polarized fluxes are expected for \object{RS~CVn}
if its emission is due to an incoherent process, since it is a high orbital
inclination binary.

On October 3, the flux detected at C band was highly polarized. The upper
limit on the flux at X band indicates a negative spectral index $\alpha_{\rm
4.9 - 8.5}$. In addition, the upper limit at L band, more than seven times
larger than the flux at C band, is compatible with a negative $\alpha_{\rm 1.4
- 4.9}$. Assuming a negative spectrum, and combined with the relatively low
radio luminosity at C band, ($L_{\rm 4.9\,GHz} =$
2.2$\times$10$^{15}$erg~s$^{-1}$~Hz$^{-1}$), ten times lower than that on June
6, the emission observed is consistent with a quiescent source. The very high
degree of polarization at C band, $\pi_{\rm c} =$ 94\%, suggests that the
circular polarization observed is likely due to a coherent emission process.
The brightness temperature at C band is $T_{\rm b} =
1.6\times10^{9}(R/$R$_{\sun})^{-2}$~K, where $R$ is the radius of the emitting
source. A brightness temperature $T_{\rm b} \gtrsim 10^{11}-10^{12}$ K can be
obtained assuming a size of the emitting region $R \lesssim 0.13-0.04$
R$_{\sun}$ at C band, respectively, in agreement with typical source sizes $R
<< R_{\star}$ for coherent emission, where $R_{\star}$ is the radius of the
star.

\section{Conclusions}\label{sec3}

We have presented multi-frequency radio observations of several RS~CVn systems
with the VLA at different epochs. The main results of our observations are
summarized as follows:

\begin{enumerate}

\item The flux spectra and the values of the degrees of circular polarization
$\pi_{\rm c}$ at different frequencies for most of the observations are
consistent with incoherent, gyrosynchrotron emission from mildly relativistic
particles. In particular, the observations at 4.9~GHz and 8.5~GHz show a clear
trend toward lower circular polarization with increasing flux. 

\item Gyrosynchrotron models predict a decrease of $\pi_{\rm c}$ when the
source becomes optically thin above a given frequency. In contrast, we found
several examples for which such a correlation is the opposite, namely
\object{HD~8357}, \object{$\sigma^{2}$ CrB}, \object{HD~8358}, \object{EI~Eri}
and \object{DM~UMa}, which show an increase of $\pi_{\rm c}$ with increasing
frequency at high frequencies.

\item We observed a helicity reversal with increasing frequency in the spectra
of three non-eclipsing systems with different spectral shapes. Two of them,
\object{EI~Eri} and \object{DM~UMa}, with a moderate flaring emission, showed
the reversal for positive spectral indexes $\alpha_{\rm 1.4 - 4.9}$ between
1.4~GHz and 4.9~GHz. On the other hand, the helicity reversal for
\object{HD~8358} took place for a quiescent spectrum with negative
$\alpha_{\rm 1.4 - 4.9}$. We found evidence for coherent emission at 1.4~GHz
in the helicity reversal of \object{HD~8358}, likely due to a plasma emission
process, as suggested by the very high $\pi_{\rm c}$ and brightness
temperature, as well as the different flux evolution of one polarization
component relative to the other. In contrast, the moderate flaring emission 
in the helicity reversal of \object{EI~Eri} and \object{DM~UMa} showed no
evidence of a different modulation in the evolution of the polarization
components at 1.4~GHz. However, the observed degrees of circular polarization,
$\pi_{\rm c} = 27-43$\%, at 1.4~GHz are too high to be accounted for by
emission from an optically thick gyrosynchrotron source. An alternative
explanation for the helicity reversal of \object{DM~UMa} on August 31, with
$\pi_{\rm c} = 5$\% at 1.4~GHz, is that the reversal is due to the transition
from an optically thick source at low frequencies to an optically thin source
at high frequencies.

\item The non-detection of circular polarization in the fluxes of 
\object{ER~Vul} and \object{RS~CVn} is in agreement with the expected trend of
decreasing degree of circular polarization with increasing orbital inclination
for incoherent emission. The sole exception is the highly circularly polarized
flux of \object{RS~CVn}, $\pi_{\rm c} = 94$\%, measured at 4.9~GHz in one
epoch. However, the mechanism invoked to explain this very high $\pi_{\rm c}$
is a coherent emission process.

\item We observed U-shaped flux spectra at high frequencies for
\object{ER~Vul}. To our knowledge, no previous observations of such U-shaped
spectra have been reported for a RS~CVn$-$type system. The low frequency part
of the spectra is consistent with gyrosynchrotron emission from an optically 
thin source. The contribution of a thermal plus a non-thermal emitting source 
is a plausible explanation for the observed shape at high frequencies. The
range of values of the brightness temperature for the flux detected at the
highest frequency is compatible with the thermal emitting source hypothesis.

\end{enumerate}

\begin{acknowledgements}

We acknowledge E. Franciosini for her useful comments and suggestions after reading through a draft version of this paper.
We acknowledge detailed and useful comments from S.~A. Drake, the referee of this paper.
J.~M.~P. and M.~R. acknowledge partial support by DGI of the Ministerio de Ciencia y Tecnolog\'{\i}a (Spain) under grant AYA2001-3092, as well as partial support by the European Regional Development Fund (ERDF/FEDER).
During this work, M.~R. has been supported by a fellowship from CIRIT (Generalitat de Catalunya, ref. 1999~FI~00199).
This research has made use of the NASA's Astrophysics Data System.

\end{acknowledgements}


\end{document}